\documentclass[fleqn,10pt]{wlscirep}
\usepackage{comment}

\title{Initial pseudo-steady state $\&$ asymptotic KPZ universality in semiconductor on polymer deposition}

\author[1,$*$]{Renan A. L. Almeida}
\author[2]{Sukarno O. Ferreira}
\author[2]{Isnard Ferraz}
\author[2,${\dag}$]{Tiago J. Oliveira}

\affil[1]{Tokyo Institute of Technology, Department of Physics, 2-12-1 Ookayama, Meguro-ku, Tokyo, 152-8551, Japan}
\affil[2]{Departamento de F\'isica, Universidade Federal de Vi\c cosa, 36570-900, Vi\c cosa, Minas Gerais, Brazil}
\affil[$*$]{lisboa.r.aa@m.titech.ac.jp}
\affil[$\dag$]{tiago@ufv.br}

\begin{abstract}

The Kardar-Parisi-Zhang (KPZ) class is a paradigmatic example of universality in nonequilibrium phenomena, but clear experimental evidences of asymptotic 2D-KPZ statistics are still very rare, and far less understanding stems from its short-time behavior. We tackle such issues by analyzing surface fluctuations of CdTe films deposited on polymeric substrates, based on a huge spatio-temporal surface sampling acquired through atomic force microscopy. A \textit{pseudo}-steady state (where average surface roughness and spatial correlations stay constant in time) is observed at initial times, persisting up to deposition of $\sim 10^{4}$ monolayers. This state results from a fine balance between roughening and smoothening, as supported by a phenomenological growth model. KPZ statistics arises at long times, thoroughly verified by universal exponents, spatial covariance and several distributions. Recent theoretical generalizations of the Family-Vicsek scaling and the emergence of log-normal distributions during interface growth are experimentally confirmed. These results confirm that high vacuum vapor deposition of CdTe constitutes a genuine 2D-KPZ system, and expand our knowledge about possible substrate-induced short-time behaviors.

\end{abstract}

\begin{document}

\flushbottom
\maketitle

\section*{Introduction}

The Kardar-Parisi-Zhang (KPZ) equation \cite{KPZ86}
\begin{equation}
\partial_t h = \nu \nabla^2 h + \frac{\lambda}{2} (\nabla h)^{2} + \sqrt{D}\eta(\textbf{x},t),
\label{eqKPZ}
\end{equation}
originally describes interface motion under conditions of no bulk conservation and exponentially fast relaxation \cite{Spohn16}. The height field $h(\textbf{x},t)$ is measured from a $d_s$-dimensional substrate at location $\textbf{x}$, with $\textbf{x} \in$ $\mathbb{R}^{d_s}$ at time $t \geq 0$. $\nu$, $\lambda$ and $D$ are phenomenological parameters, physically representing the surface tension, the excess of velocity in the growth, and the amplitude of a space-time white noise $\eta$, respectively.

Although posed 30 years ago, outstanding advances on the understanding of the KPZ class have been made quite recently. Following seminal works on multiple-meaning stochastic models \cite{Johansson00, Spohn00}, long-awaited analytical solutions \cite{Spohn16,Sasamoto16}, experiments \cite{Takeuchi11, HTakeuchi15} and numerical simulations \cite{HTakeuchi15, Alves11, Tiago12} came out to confirm that \textit{asymptotic} 1D-KPZ height distributions (HDs) are related to statistics of the largest eigenvalues of random matrices \cite{TW94}, while spatial covariances are dictated by the \textit{time}-correlation of Airy processes \cite{Covariance}. Noteworthy, both HDs and covariances exhibit sensibility to initial conditions (ICs) \cite{Spohn00, Covariance}, splitting the KPZ class into subclasses according to the ICs. This unanticipated feature was recently observed also in models for nonlinear molecular beam epitaxy class \cite{Ismael16}. Similar scenario has been found for the 2D-KPZ case, based on numerical simulations \cite{Healy12, Tiago13, Ismael14}, although no analytical result is known for 2D-KPZ HDs and covariances and the existing theoretical approaches \cite{Canet10,Kloss12} for the scaling exponents disagree with numerical outcomes. In such arid landscape, the rare reliable experimental evidences of 2D-KPZ universality \cite{Cuerno00, Renan14, Healy14, Renan15} turns to be precious achievements.

The short-time roughening of systems exhibiting asymptotic KPZ scaling is also rich in behavior. For example, a transient scaling in the Edwards-Wilkinson \cite{barabasi} (EW) class might appear whenever $\lambda$ is ``small'' when compared to $\nu$ and $D$. On the other hand, if these parameters satisfy the condition $D \gg (\nu, \lambda)$, then a transient Random Deposition (RD) scaling  might take place. In both cases, a crossover to KPZ dynamics occurs at a characteristic time $t_c$. The EW-KPZ and RD-KPZ crossovers were intensively studied in competitive growth models \cite{Tiago13EWKPZ,Tiago13RDKPZ,Alice15}, but are also related to the edge statistics of fermionic lattices at high-temperatures \cite{LeDoussal16} and to synchronization problems in parallel computation \cite{Alice15}. Besides a recent observation of RD-KPZ crossover in CdTe/Si(100) films deposited at $T = 200\,^{\circ}\mathrm{C}$ \cite{Renan15}, experimental studies showing crossovers to asymptotic KPZ scaling are mostly lacking, even though should be quite expected.

In this contribution, we show experimentally that a new type of short-time regime, differing from EW and RD, may take place during the growth of a semiconductor film on a polymeric substrate. Such system has a very important technological application in the fabrication of flexible solar cells \cite{Romeo06}. By analyzing surface fluctuations of CdTe films deposited on polyimide (Kapton) substrates, we reveal that the observed transient regime is characterized by Gaussian statistics, a constant roughness in time and \textit{no} spreading of correlations through the system. Since these last characteristics are hallmarks of ``saturated'' interfaces, we will refer to that regime as a \textit{pseudo}-steady state (PSS). After a long characteristic time $t_c$, corresponding to the deposition of $\sim 10^4$ monolayers of CdTe, correlations and height fluctuations start developing, yielding an asymptotic growth in the KPZ class. We show in fine experimental details properties of 2D-KPZ universality, by independently measuring several scaling exponents, universal distributions related to the height, roughness and extremal height fluctuations, as well as the rescaled spatial covariance. Experimental evidence of log-normal distributions in surface growth dynamics is also given.

\section*{Experimental methods}

We performed the experiment growing CdTe films on polymeric substrates (Kapton, Dupont) by hot wall deposition technique in high-vacuum ($\sim 10^{-7}$ Torr). This growth technique has been chosen for its simplicity and for being demonstrated to yield high quality CdTe films with properties similar to films produced by molecular beam epitaxy \cite{Sukarno03,Suela10}. Substrate temperature was set to $T = 150\,^{\circ}\mathrm{C}$, and source temperature to $510\,^{\circ}\mathrm{C}$, yielding a growth rate $F = 14.0(3)$ $nm/$min. Different films were grown for several times varying from $7.5$ min to $5760$ min. Before the deposition, substrates were annealed (at $T = 150\,^{\circ}\mathrm{C}$) inside the growth chamber for $15$ min in order to release weakly bounded impurities and guarantee thermal homogeneity since the beginning of the deposition process. For each time a completely independent film was grown and its morphology was characterized by atomic force microscopy, using an NTEGRA-Prima SPM (NT-MDT). We collected about $\sim 10^{7}$ spatial points in total, in up to 10 different scanned regions for each film (time). Experimental data were analyzed by our own computational algorithms. The results were checked to be statistically independent of the AFM tip, lateral scan size ($L$) and operational mode (contact or tapping). In the following we show data from images taken at $L = 10$ $\mu$m, with 512 $\times$ 512 pixels, and contact mode. X-ray diffraction, with a D8-Discover diffractometer (BRUKER), was used to characterize the crystalline structure of the grown films.


\section*{Results and discussions}

\subsection*{Pseudo-steady state crossover to KPZ}

\subsubsection*{Experiments}

\begin{figure}[!t]
\centering
\includegraphics[width=17.5cm]{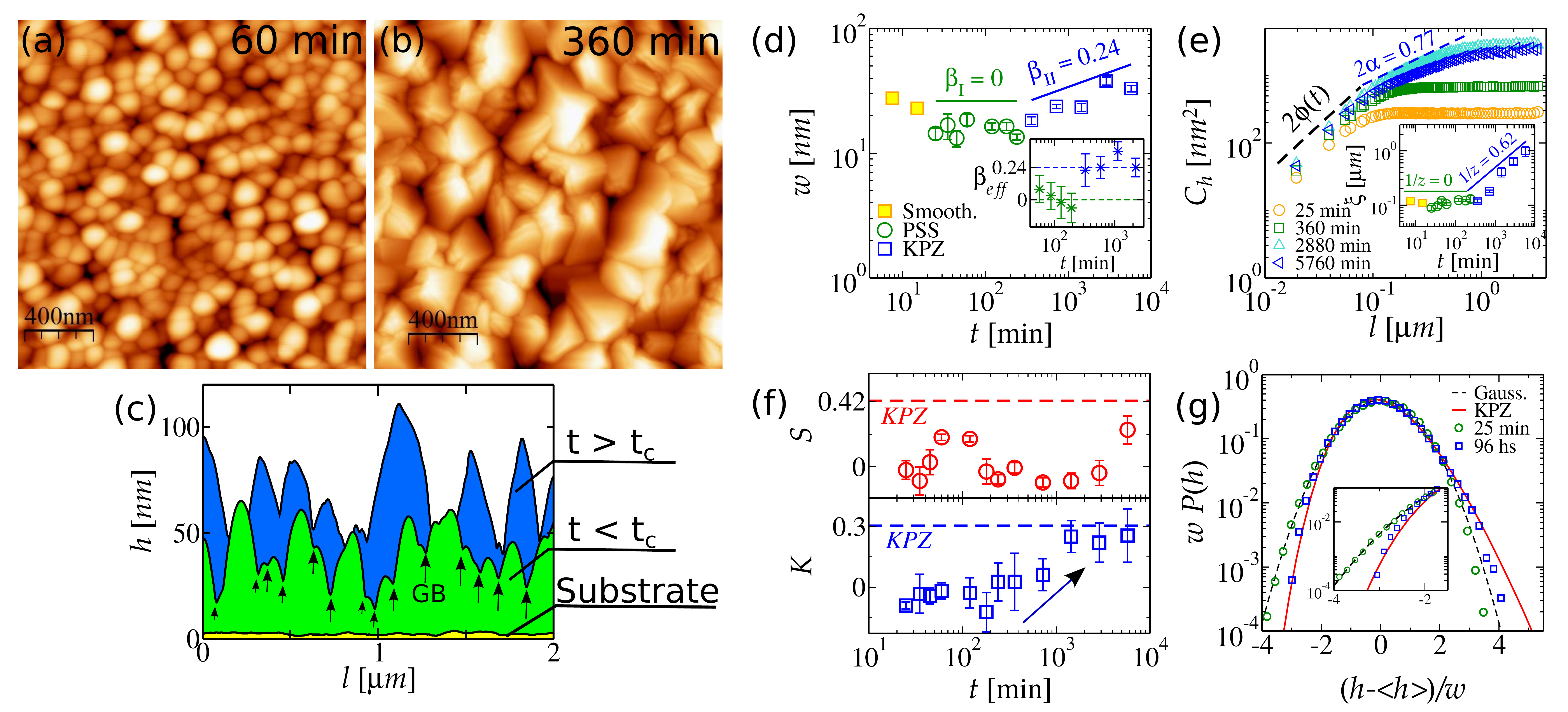}
\caption{AFM images of CdTe surfaces for films grown at (a) 60 min (PSS regime) and (b) 360 min (beginning of the KPZ scaling). Typical 1D profiles are sketched out in (c). Black arrows indicate grain boundary (GB) sites considered in MC simulations. (d) Global CdTe surface width ($w$) as function of $t$. Different colored symbols and solid lines indicate different regimes. Inset shows the effective $\beta$ exponent. (e) Height difference correlation function ($C_h$) versus $l$ for different deposition times. Dashed lines are guides to eyes with slopes 1.2 and 0.77 (KPZ)\cite{Healy12}. Inset shows the temporal variation of the correlation length $\xi$. (f) Temporal evolutions of HDs' skewness ($S$) and kurtosis ($K$). The dashed lines indicate the universal KPZ values. (g) Rescaled HDs (with null mean and unity variance) for CdTe films (symbols), compared with Gaussian (dashed) and KPZ (solid line)
HDs. Inset shows the same data highlighting the left tails.}
\label{fig1}
\end{figure}

Surface morphologies of grown samples can be visually distinguished in two sets [see Fig. \ref{fig1}]. For times shorter than $\approx 300$ min, the surface is constituted by a plethora of rounded, compacted grains, with average size $l_g \approx 100$ $nm$ [Fig. \ref{fig1}(a)]. X-Ray diffraction measurements reveal a polycrystalline structure with crystallites in a wide spectrum of orientations [Suppl. Inf. 1]. At longer times, the (111) growth orientation dominates, giving rise to grains ($l_g \approx 300$ $nm$) with approximately pyramidal shapes. [Fig. \ref{fig1}(b)]. Interesting, the asymptotic pyramidal morphology resembles the typical KPZ-CdTe patterns observed during the growth of CdTe on Si(001) substrates \cite{Renan14,Renan15}. One-dimensional profiles of these two distinct regimes [Fig. \ref{fig1}(c)] give additional support for the description above, where a profile for the initially rough Kapton substrate (which has width $w_0\approx 6$ $nm$, and correlation length $\xi_0 \approx 0.3$ $\mu m$) is also shown.

Surface width $w(L,t)$, defined as the standard deviation of the height field, also exhibits different temporal regimes [Fig. \ref{fig1}(d)]. For very short times $t \in [7.5, 25)$ min, $w$ decreases from 28(3) $nm$ to 14(2) $nm$ - numbers into parenthesis represent uncertainties in the last digits. This time interval is related to an interface smoothening mechanism, which will be explained below. Within the interval $[25, 300)$ min the width does not grow, so that applying the usual $w \sim t^{\beta}$ scaling \cite{FV85} we find $\beta_I = -0.01(7)$, where $\beta$ is the growth exponent. This small exponent can also be interpreted as a logarithmic behavior of the roughness [$w(t) \sim \sqrt{\ln(t)}$], a fingerprint of EW class in 2D \cite{barabasi}. Finally, for $t \gtrsim 300$ min, usual power-law scaling arises with $\beta_{II} = 0.24(7)$. Such value provides an evidence of KPZ scaling (Tab. \ref{t1}), although it also encompasses (within the error bar) the exponents for the linear ($\beta_l = 0.25$) and nonlinear ($\beta_n=0.20$) molecular beam epitaxy classes \cite{barabasi}. Defining $\beta_{eff}$ as the slope of five consecutive points in the $\log w \times \log t$ plot [Fig. \ref{fig1}d], we find compelling evidence that $\beta_{eff}$ takes two distinct values separated by a characteristic time $t_c \approx 300$ min [inset of Fig. \ref{fig1}(d)]. Namely, one has $\beta_{eff} \approx 0$ for $t < t_c$ (the smoothening regime is not considered) and $\beta_{eff} \approx 0.24$ otherwise.

Now we turn to characteristic length scales, which are investigated by the (equal-time) height difference correlation function $C_h \equiv \langle[h(\textbf{x + l}, t) - h(\textbf{x},t)]^2 \rangle$. Brackets represent averages over different AFM images and positions $\textbf{x}$, and $l$ varies in the interval $[0, L]$. Accounting for the grains at surface (with average lateral size $l_g$), $C_h(l)$ is expected to scale as \cite{Tiago}:

\begin{equation}
C_h(l) \sim \left\{ \begin{array}{ll}
l^{2\phi} & \textrm{for} \quad l \ll l_g,\\
l^{2\alpha} & \textrm{for} \quad l_g \ll l \ll \xi,\\
const. & \textrm{for} \quad \xi \ll l,
\end{array} \right.
\label{corr}
\end{equation}
where $\phi$ is a nonuniversal exponent related to the grain shape \cite{Tiago} and $\alpha$ is the (universal) roughness exponent \cite{FV85}. Figure \ref{fig1}(e) shows $C_h(l)$ calculated for CdTe surfaces at different times. For $t \lesssim 720$ min, solely the scaling for $l \ll l_g$ is observed, from which one obtains a time-dependent exponent $\phi(t) \in [0.6,0.8]$. At longer times, the correlation length turns out to be larger than $l_g$, so that from the second scaling regime we measure $\alpha = 0.37(3)$ - in striking agreement with the best known estimate for the 2D-KPZ class [Tab. \ref{t1}]. We remind that roughness exponents for EW, linear and nonlinear molecular beam epitaxy classes are, respectively, $\alpha_{EW} = 0$, $\alpha_l = 1$ and $\alpha_n=2/3$. Hence, $\alpha = 0.37(3)$ is our first strong evidence of KPZ scaling. As an aside, we point out that a clear scaling regime $l_g \ll l \ll \xi$ has not been observed for CdTe/Si(100) system \cite{Renan14,Renan15}, possibly due to the short growth times analyzed there, as is also the case of the samples grown up to 720 min here. This is one of the reasons why a consistent universal $\alpha$ exponent is rarely observed in real grained surfaces.
 
At the beginning of the growth process, the correlation length is arguably of the same order of grains' size $\xi \simeq l_g$. Thence, the single crossover in $C_h$ curves provides an estimate for $\xi$. For long times, $\xi$ is associated with the second crossover. The insertion in Fig. \ref{fig1}(e) shows the evolution of $\xi$, from where one sees that during the interval in which the width does not grow, $t \in [25$ min$, t_c]$, correlations also \textit{do not}. Interesting, these properties are presented by finite-size interfaces at their steady states, where both $w$ and $\xi$ remain constant in time. Indeed, by assuming a power-law scaling like $\xi \sim t^{1/z}$, we find $1/z = -0.04(3)$ for $t < t_c$. On the other hand, in the asymptotic regime $t \gg t_c$ correlations start spreading through the system with a characteristic exponent $1/z \approx 0.63$, in the same way as stochastic models belonging to the KPZ class do [Tab. \ref{t1}]. Similar results (not shown) are obtained by estimating $\xi$ from the first zeros (or first minima) of the slope-slope correlation function, as done, e. g., in \cite{Renan14,Renan15}.

These results demonstrate that the first regime is a \textit{pseudo}-steady state (PSS), not described by the EW equation, for which $1/z = 0.5$ is expected \cite{barabasi}, while $1/z \approx 0$ is found here. On the other hand, the asymptotic regime is KPZ, as strongly suggested by the three independently measured exponents $\alpha$, $\beta$ and $1/z$. In addition, the Galilean invariance relation $\alpha + z = 2$ \cite{KPZ86} is quite well satisfied, with $\alpha + z \approx 1.96$. Table \ref{t1} summarizes the exponents obtained for each regime at side with the best known numerical estimates for 2D-KPZ class.

Last, we point that for $t < t_c$ height distributions (HDs) are Gaussian, while for $t \gg t_c$ they reasonably agree with the universal KPZ one. Figure \ref{fig1}(f) shows temporal evolutions of the skewness $S \equiv \langle h^3 \rangle_c/\langle h^2 \rangle_c^{3/2}$ and kurtosis $K \equiv \langle h^4 \rangle_c/\langle h^2 \rangle_c^2$ (here $\langle X^n \rangle_c$ will denotes the $n^{th}$ cumulant of a random variable $X$).  Within the PSS regime, one sees that $S \approx 0$ and $K\approx 0$. A clear crossover is observed in $K$, which approaches to the KPZ value $K_{KPZ} \approx 0.34$ \cite{Healy12, Tiago13} at long times. The skewness, on the other hand, seems to start converging latter and is still smaller than the KPZ one even for the longest time analyzed [Tab. \ref{t1}]. Unfortunately, with the few data points available in the convergence region, we are not able to extrapolate $S$. Anyhow, Fig. \ref{fig1}(g) gives additional evidence of a Gaussian-to-KPZ crossover in the rescaled HDs, $P(h)$, as seen by the good collapse between the respective distributions in each regime. The agreement with KPZ HD is particularly clear in the left tails [inset of Fig. \ref{fig1}(g)] and at the peaks. Some deviations, however, are observed in the right tail, due to the skewness a bit smaller than the \textit{asymptotic} KPZ one.

\begin{table}[t]
\centering
\begin{tabular}{|l|l|l|l|}
\hline
\textbf{} & $t < t_c$ & $t \gg t_c$ & \textbf{KPZ models} \\
\hline
$\alpha$ & not determined & 0.37(4) & 0.3869(4) \cite{Pagnani16} \\
\hline 
$\beta$ & -0.01(7) & 0.24(7) & 0.241(1) \cite{Odor16} \\
\hline
$1/z$ & -0.04(3) & $\approx$ 0.63  & 0.623(3) \cite{Pagnani16,Odor16} \\
\hline
$\alpha + z$ & not determined & $\approx$ 1.96 & 2 \cite{KPZ86} \\
\hline
$S$ - [$P(h)$] & 0.03(6) & 0.24(9) & 0.42(1) \cite{Healy12, Tiago13} \\
\hline
$K$ - [$P(h)$] & -0.06(8) & 0.3(1) &  0.34(1) \cite{Healy12, Tiago13} \\
\hline
\end{tabular}
\caption{Scaling exponents ($\alpha$, $\beta$ and $1/z$) and HDs' cumulant ratios ($S$ and $K$) for CdTe/Kapton surfaces at short ($2^{nd}$) and long ($3^{rd}$ column) deposition times. The rightmost column shows the best known estimates for those quantities obtained from numerical simulations of 2D-KPZ models. For $t<t_c$, the experimental values of $S$ and $K$ are averages, while the ones for $t\gg t_c$ are those for the longest times investigated.}
\label{t1} 
\end{table}

\subsubsection*{Modeling}

In order to better comprehend the origins of the distinct regimes in the roughness scaling, relevant microscopic ingredients of the CdTe growth shall be clarified. At the very beginning of the deposition process, one observes that the roughness fast increases from $w_{0} \approx 6$ $nm$ (Kapton substrate) to $w = 28(3)$ $nm$ (for $t=7.5$ min), as inferred from Fig. \ref{fig1}(d). First, deposition temperature is relatively low ($T=150 \,^{\circ}\mathrm{C}$), so that diffusion of adsorbed species is slow, leading to the nucleation of a large number of CdTe islands at the submonolayer regime. Second, it seems to exist a strong aversion of CdTe molecules into wetting the Kapton surface, inducing the formation of 3D islands (grains), according to the Volmer-Weber growth mode. Thereby, islands with small lateral sizes ($l_g$) and large heights ($h_g$) are initially formed, which may explain the fast $w$ increasing.

To understand the subsequent smoothening (the decreasing $w$) and PSS regimes, the key points are: $i)$ CdTe layers have been observed to develop a strong texture in the (111) direction \cite{Renan14, Renan15}, implying that (111)-grains grow faster than the ones with other crystallographic orientations; and $ii)$ the amorphous nature of the Kapton substrate leads to the formation of CdTe grains in a wide spectrum of crystallographic orientations, so that (111)-grains shall take a long time to dominate the surface. This is in contrast with the situation on (crystalline) Si substrates, where a majority of (111)-grains is observed since short times [Suppl. Inf. 1]. Keeping these informations in mind, CdTe/Kapton roughness scaling can be explained as follows. When initial grains collide, forming a film that completely covers the substrate, some large height differences at surface obviously disappear, leading to the fast smoothening observed for $t \in [7.5, 25)$ min. In the subsequent multilayer regime, we may expect that the CdTe growth dynamics proceed as explained in \cite{Renan15}: defect sites at and around the grain boundaries (GBs) between two or more (collided) grains inhibit mass transport of adsorbed species between them, preventing the spreading of correlations on the surface. For instance, in CdTe/Si(100) system (deposited at $T=150 \,^{\circ}\mathrm{C}$) such condition led to a Random Deposition growth (for which $\beta=1/2$ and $1/z=0$) at short times \cite{Renan15}, with grain peaks evolving in a uncorrelated way. For the present system, one could expect a similar behavior, and indeed we find $1/z \approx 0$ for $t < t_c$. However, there exists a key distinction here. Because of the fact in $ii)$, grains with large height [mostly the (111) ones] are considerably surrounded by smaller [mostly non-(111)] ones up to relatively long times. When the low height [non-(111)] grains are covered up by higher (111) ones, allowing these ones to collide, a smoothening mechanism similar to that of the very initial growth times happens. In short, while deposition tends to yield an increasing roughness, collisions of (111) grains after covering other ones lead to a smoothening. Very interesting, during a long time interval ($t \in [25,300]$ min) both mechanisms compensate each other, so that $w$ (and $\xi$) remains constant, giving rise to the PSS regime. It is worthy mentioning that the smoothening operative here - driven by grain collisions - is quite different from that studied in stochastic growth models deposited on rough substrates \cite{TiagoBA}.

At very long growth times, surface becomes dominated by (111) crystallites - the probability of finding a (111) grain in the film becomes $\gtrsim 90\%$ [Suppl. Inf. 1] -,  so that the smoothening effect becomes negligible. Relaxation of defects at GBs eventually occurs and coalescences and spreading of correlations turn out to be the relevant mechanisms for the subsequent dynamics. Coalescence/packing of (111) grains is already known to yield a velocity excess in CdTe growth, leading to the KPZ scaling \cite{Renan14, Renan15}.

The reasoning above is corroborated by a discrete 1D model, proposed in Ref. \cite{Renan15} to explain the short-time coalescence dynamics of CdTe/Si(001) films. Since our aim here is to demonstrate the role of a large initial roughness and GB defects to yield the PSS regime, we use (1D) height profiles extracted from (2D) CdTe images (at $t = 60$ min) as the initial condition. Moreover, to mimic the experimental situation, where the CdTe lattice constant is $\sim 1$ $nm$, we assume the same for the simulations and rescale the profiles accordingly (e. g., a profile of 2 $\mu m$ shall correspond to $L \sim 2000$ sites, which is obtained by enlarging the original 512 pixel profile by a factor 4). For special sites, mimicking the grain boundaries (GBs), an energy barrier $E_{GB}=0.10$ eV is locally assigned, in order to hamper diffusion of particles towards that region [for sake of simplicity we set GB sites at the local minima at the profiles - see Fig. \ref{fig1}(c)]. It is assumed that diffusion and aggregation occur much faster than adsorption (desorption is not considered), so that each particle permanently aggregates before the arrival of the next one at surface. Thus, for each event of deposition, the following rule holds: a particle, randomly deposited on the surface, diffuses until finding a site $i$ satisfying the restricted solid-on-solid condition $|h_i-h_{i\pm 1}| < 1$ ($h_i$ is the height of site $i$, measured from the substrate), where it permanently aggregates. Diffusion towards a GB site occurs with probability $P_d = e^{-E_{GB}/k_B T}$, where $k_B$ is the Boltzmann constant, while diffusion to regular sites occurs with unity probability ($P_d = 1$). Relaxation of defects around GBs is activated through deposition and rearrangement of particles. Then, in the model, whenever a particle is deposited at a given GB site, it might become a regular site with probability $P_R = e^{-E_{R}/k_B T}$, where we set $E_{R}=0.30$ eV. The time scale (``min'') is defined so that the deposition rate is given by 14 ML/``min'', similarly to the experiment.

\begin{figure}[!t]
\centering
\includegraphics[width=11cm]{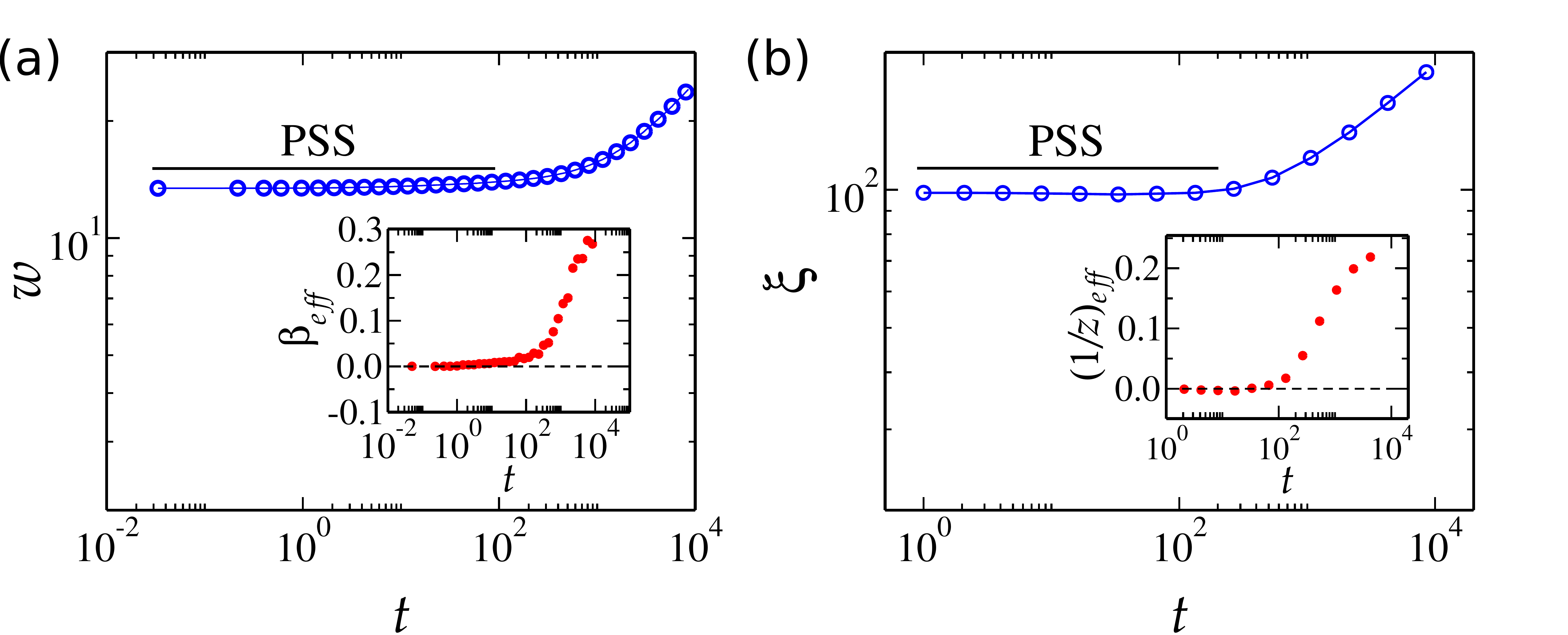}
\caption{Temporal evolutions of (a) interface width $w$ and (b) correlation length $\xi$ for the 1D model. Insertions show the effective (a) growth $\beta_{eff}$ and (b) inverse dynamic $1/z_{eff}$ exponents as functions of time.}
\label{fig2}
\end{figure}

A clear regime of constant roughness is observed at initial times of the growth model, being replaced by a scaling regime at longer times [Fig. \ref{fig2}(a)]. The effective growth exponents confirm both behaviors, since $\beta_{eff} \approx 0$ is found for $t < t_c$ and $\beta_{eff} \neq 0$ for $t \gg t_c$ [inset of Fig. \ref{fig2}(a)]. Coincidentally, there exist even quantitative agreement between both $w$ ($\approx 13$ at short times) and $t_c$ ($\approx 300$ ``min''), with the experiment: $w \approx 15$ nm (in PSS regime) and $t_c \approx 300$ min [Fig. \ref{fig1}(d)]. The correlation length parallel to substrate ($\xi$) is shown in Fig. \ref{fig2}(b) as a function of time, with the respective effective (inverse) dynamic exponents ($1/z_{eff}$) displayed in the insertion. The almost constant $\xi$ at short times ($t<t_c$), yielding $1/z_{eff} \approx 0$, confirms the existence of a PSS regime in the model. This strongly suggests that the interplay of defects at GBs (inhibiting intergrain diffusion and grain coalescence, as well as the spreading of correlations at surface) and an initial condition with grains of very different heights may be the origin of the PSS in CdTe/Kapton, as explained above.

Despite the success of this model for explaining the short-time behavior of CdTe growth, as also observed in \cite{Renan15}, we stress that \textit{no} agreement with experiments is expected for $t \gg t_c$, since the model clearly does not have an asymptotic scaling in KPZ class. Moreover, it is 1D and cannot explain the asymptotic dynamics of a 2D system. For this matter, a much more complex off-lattice model is required taking into account the existence of grains with different orientations and growth velocities, as well as the effect of grain packing and so on.


\subsection*{Generalized Family-Vicsek scaling and log-normal distributions}

From this section we explore CdTe surfaces as an experimental testbed for additional scaling and distributions theoretically predicted for the KPZ class and, conversely, obtain further confirmation of the results presented so far. Let us start with a recent generalization of the Family-Vicsek (FV) \cite{FV85} scaling \textit{ansatz}, which has not been verified experimentally yet. According to Ref. \cite{Ismael15}, the squared local width, $w_2(l,t)$ - calculated in square windows of lateral size $l \ll L$ that span the whole surface - is a fluctuating variable whose $n^{th}$ cumulant of its distribution [$P(w_2)$] behaves as:

\begin{equation}
\langle (w_2)^n \rangle_c = l^{2n\alpha}f_n(t/l^z),
\label{ismael}
\end{equation}
where $f_n(u) \sim u^{\gamma_n}$ for $u \ll 1$, and $f_n(u) \sim const$ for $u \gg 1$. The $\gamma_n$ exponents are given by $\gamma_n = {2n\beta + [(n-1)d_s]/z}$, and their values for 2D-KPZ class, calculated from the best known estimates of $\beta$ and $z$, are summarized in Tab. \ref{t2}. Note that for $n=1$, the classical FV scaling is recovered.

\begin{figure}[t]
\centering
\includegraphics[width=17.5cm]{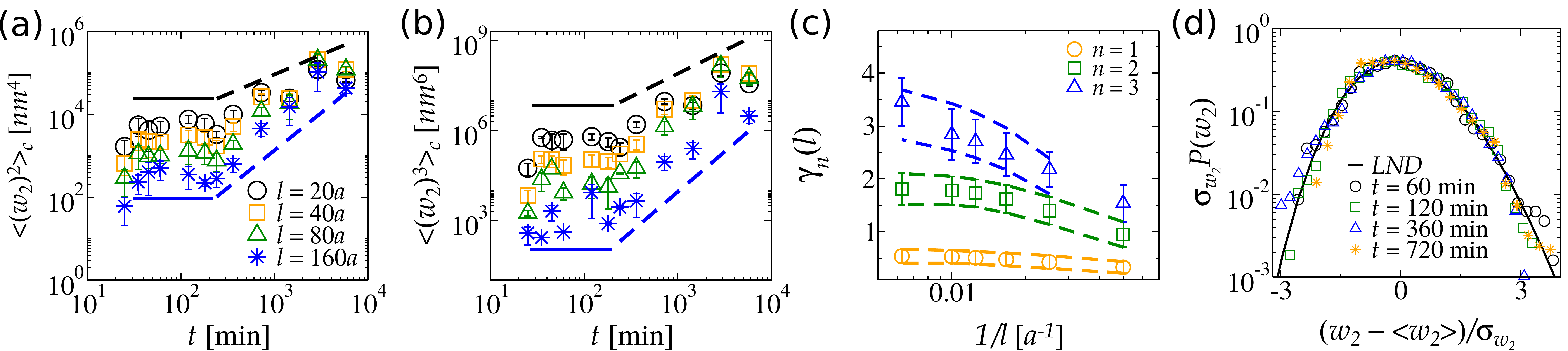}
\caption{Temporal evolution of the second (a) and third (b) cumulants of width distributions, $P(w_2)$, calculated for windows of different lateral sizes $l$. Solid and dashed lines are guide to eyes indicating the PSS and KPZ regime, respectively. Top (black) and bottom (blue) dashed lines have the slope of the curves for $l = 20a$ and $l = 160a$, respectively. (c) Exponents $\gamma_n(l)$ as a function of $1/l$. (d) Rescaled width distributions for CdTe surfaces (symbols), at several growth times and calculated for $l = 80a$, along with the LND (solid line). Here, $\sigma_{w_2} \equiv \sqrt{\langle (w_2)^2 \rangle_c}$.}
\label{fig3}
\end{figure}

Bearing in mind that $\gamma_n$ exponents are related to the regime for which $\xi \ll l$, and relied upon our estimate for $\xi$ made in Fig. \ref{fig1}(e), we conclude that a suitable interval for performing experimental measurements is $l \in [20, 160]a$, with $a \equiv 10$ $\mu m$/512 being the pixel size. The upper limit ($160a \ll L$) is chosen to guarantee a good statistics. Figures \ref{fig3}(a) and \ref{fig3}(b) display, respectively, the variation of the second and third cumulants of $P(w_2)$ in time. The initial almost constant behaviors observed, consistent with $\beta \approx 0$ and $1/z \approx 0$, give further evidence of the existence of a PSS regime. Algebraic scaling arise at long times, from which the exponents shown in Fig. \ref{fig3}(c) are measured. Hydrodynamic limit demands finding $\gamma_n$ after taking the limits $l (\leqslant L) \rightarrow \infty$ and $t \rightarrow \infty$. Since the last requirement is not of easy experimental implementation, we analyze \textit{effective} $\gamma_n$'s for different $l$'s and extrapolate them to $l \rightarrow \infty$, as done in Fig. \ref{fig3}(c). Exponents obtained from such procedure are summarized in Tab. \ref{t2}, with all values consistently in agreement, within the error bars, with those expected for the KPZ class.

A more complete characterization of $w_2$ fluctuations is set down by the (full) width distributions, $P(w_2)$. For a given surface, as the window size $l$ approaches to $L$, with $\xi \ll l$, $P(w_2)$ approaches to a Dirac delta function (because the number of windows converge to unity). More interesting, however, is that $P(w_2)$ converges first to a log-normal distribution (LND) before reaching the delta. As explained in Ref. \cite{Ismael15}, the emergence of LNDs is a general feature of low-correlated growth (when $\xi \ll l$) that does not depend on the universality class. We experimentally test this conjecture by comparing rescaled (to zero mean and unity variance) $P(w_2)$ for CdTe interfaces, for large $l$ and small $t$, with the LND [Fig. \ref{fig3}(d)]. Experimental distributions collapses nicely onto LND at their peaks and tails, for \textit{both} PSS and KPZ regimes, confirming the role of LND in surface growth context. A more detailed analysis of this subject is presented in Suppl. Inf. 2.


\subsection*{Stationary fluctuations}

\begin{figure}[!h]
\centering
\includegraphics[width=17.5cm]{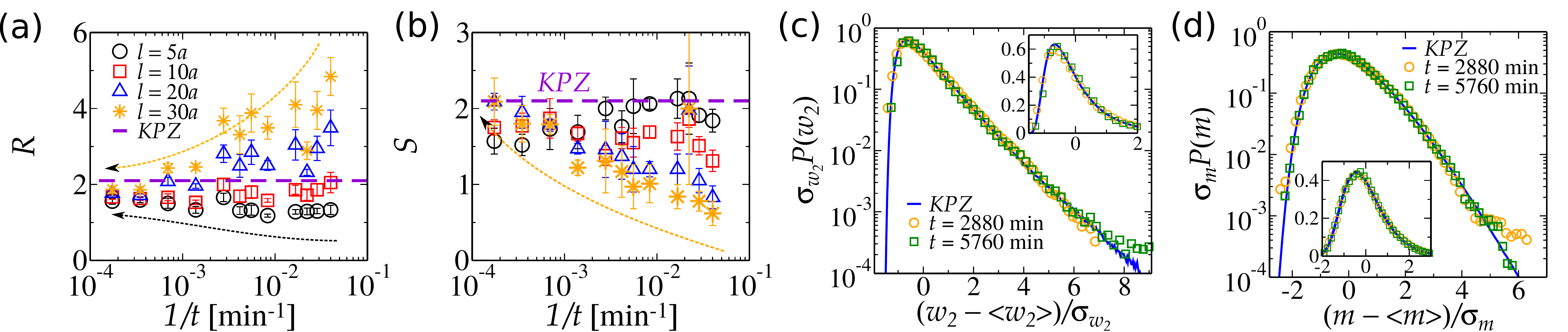}
\caption{(a) Inverse of the variance coefficient $R$ and (b) skewness $S$ as function of $1/t$ for (stationary) width distributions, for several window sizes $l$. Rescaled (c) width [$P(w_2)$] and (d) maximal height [$P(m)$] distributions, for CdTe (symbols) and 2D-KPZ models (solid blue lines). Insets show the same data in log-linear scale. Here, $\sigma_{X} \equiv \sqrt{\langle X^2 \rangle_c}$ is the standard deviation of $X = w_2,m$.}
\label{fig4}
\end{figure}

While width fluctuations for $\xi \ll l \ll L$ are given by log-normal distributions regardless the universality class of the system, in the stationary limit $\xi \gg l$ different and universal pdf's, $P(w_2)$, are expected for each universality class. Width distributions for the stationary regime were analytically calculated in the 90's for 1D linear growth models \cite{Foltin94, Plischke94} assuming periodic boundary conditions (PBC). Subsequent work \cite{Antal02} called attention for the importance of considering window boundary conditions (WBC), which parallels the experimental way of obtaining $P(w_2)$. More challenging, the 2D distributions for WBC have been numerically investigated \cite{Healy14,Ismael15,Fabio15,Iuri15}, and experimentally used to confirm the universality of CdTe/Si(001) \cite{Renan14, Renan15}, oligomer \cite{Healy14} and oxide \cite{Iuri15} growing films. The successful agreement between numerical and experimental width distributions in 2D motivated their analysis also for 1D KPZ interfaces of nematic liquid-crystal system \cite{HTakeuchi15}. The systematic study of these distributions \cite{Ismael15} showed that the best way to verify their asymptotic universality is performing successive extrapolations of their cumulant ratios for $t \rightarrow \infty$ and then for $l \rightarrow \infty$. Although we are not able to implement this procedure here, due to statistical fluctuations in the data [Fig. \ref{fig4}], we find that for large $t$ and $l$ the cumulant ratios tend to approximate to the KPZ ones. A relaxed methodology relies upon guaranteeing that cumulants ratios converge to \textit{plateaus regions} at sufficiently long times, with $l \lesssim 0.3\xi$ being a safe limit over which experimentalists should prop up \cite{Fabio15}.

For CdTe surfaces, one has $\xi \gtrsim 90a$ for $t \geq 2880$ min, hence $l \leq 30a$ is considered in our analyses. In Fig. \ref{fig4}(a), the ratio $R \equiv \langle w_2 \rangle_c/\langle(w_2)^2\rangle_c^{1/2}$ (the inverse of the coefficient of variance) is plotted as function of $1/t$ for several $l$'s. Indeed, at large $t$ and $l$, experimental data systematically approach to the $R$ value expected for the 2D-KPZ class. Analogous behavior is observed for $S$ [Fig. \ref{fig4}(b)] and $K$ ratios [Suppl. Inf. 2]. The values for $t = 5760$ min and $l = 30a$, the closest ones to the asymptotics, present a remarkable agreement with the universal KPZ ratios [see Tab. \ref{t2}]. Further confirmation of such agreement is provided by the collapse of rescaled width distributions for CdTe surfaces onto the universal KPZ curve at both tails and at the peak [Fig. \ref{fig4}(c)].

\begin{table}[!t]
\centering
\begin{tabular}{|l|l||l|l||l|l|}
\hline
\textbf{CdTe}-$\gamma_n$ & \textbf{KPZ}-$\gamma_n$ & \textbf{CdTe}-[$P(w_2)$] & \textbf{KPZ}-[$P(w_2)$] & \textbf{CdTe}-[$P(m)$] & \textbf{KPZ}-[$P(m)$] \\
\hline
$\gamma_1 = 0.57(12)$ & $0.4830(30)$ & $R = 1.87(8)$ & $2.05(5)$  &$R = 7.3(5)$ & $7.3(4)$\\
\hline
$\gamma_2 = 1.96(29)$ & $2.214(15)$ & $S = 2.1(3)$ & $2.04(4)$ & $S = 0.83(9)$ & $0.84(2)$\\
\hline
$\gamma_3 = 3.60(52)$ & $3.946(27)$ & $K = 7.6(2.4)$ & $7.3(3)$ & $K = 1.19(34)$ & $1.14(5)$\\
\hline
\end{tabular}
\caption{(Left) Asymptotic $\gamma_n$ exponents from generalized FV scaling relation \ref{ismael}. The KPZ values were calculated using the estimates of $\alpha$ and $\beta$ in \cite{Pagnani16, Odor16}. Cumulant ratios $R$, $S$ and $K$ for stationary width (middle) and maximal height (right) distributions calculated with WBCs. All KPZ values for the cumulants were extracted from Ref. \cite{Ismael15}.}
\label{t2} 
\end{table}

A quite similar calculation can be made for evaluating the stationary distributions, $P(m)$, of \textit{local} extreme heights $m$, where $m = h^* - \langle h \rangle$, and $h^*$ is either the maximal or minimal value of $h$ inside a window of size $l \ll \xi$. Extreme statistics plays a relevant role in several areas of knowledge and are usually related to catastrophic events (see, e.g., \cite{Fortin15} for a recent review). The scaling of the \textit{global} $m$ and its distribution have been investigated in a number of works \cite{Raychaudhuri00, Lee05, Schehr06, Gyorgyi07, Tiago08}, while the universality of the \textit{local} distribution [$P(m)$, with WBC] has been numerically demonstrated more recently in several numerical and experimental studies \cite{Renan14,Healy14,Renan15,Iuri15,HTakeuchi15,Ismael15}. Following the same recipe used for calculating $P(w_2)$, we determine $P(m)$ (for \textit{maximal} heights) and its related cumulant ratios $(R,S,K)$, whose convergence with $t$ and $l$ is presented in Suppl. Inf. 2. Their asymptotic values, displayed in Tab. \ref{t2}, are in striking agreement with the KPZ ones. A nice collapse of rescaled distributions $P(m)$ for CdTe surfaces onto the KPZ curve is observed around three orders of magnitude from the peak [Fig. \ref{fig4}(d)]. Altogether, these results for stationary distributions give a quite compelling confirmation of KPZ universality in the growth of CdTe films on Kapton substrates.

\subsection*{KPZ ansatz and spatial covariance}

The asymptotic evolution of the (1-point) height of a growing KPZ interface can be summarized in the expression \cite{KM92}:
\begin{equation}
 h(t) \simeq v_\infty t + s_{\lambda}(\theta t)^\beta\chi,
\label{eqansatz}
\end{equation}
where $v_\infty$ (the asymptotic growth velocity), $s_{\lambda}$ (the signal of $\lambda$), and $\theta \equiv (A^{1/\alpha}\lambda)$ are system-dependent parameters. $\chi$ is a random variable distributed according to universal (height) distributions $P(\chi)$, which depend on the substrate dimension \cite{HTakeuchi15} and initial conditions \cite{Spohn00}. The so-called ``KPZ ansatz'' (eq. \ref{eqansatz}) can also describe the dynamics of growing interfaces belonging to other universality classes, provided that $\theta$ is suitably redefined \cite{Ismael16}. 

Following the procedure introduced in Ref. \cite{Healy14}, we may estimate $\theta$, $A$ and $\lambda$ for the CdTe/Kapton surfaces by using the (2-point) spatial covariance $C_s$, defined as:
\begin{equation}
 C_s(l,t) = \langle[h(\textbf{x + l}, t)h(\textbf{x},t)] \rangle - \langle h\rangle^2 \simeq (\theta t)^{2\beta}g\left[ \frac{A_h}{2}l^{2\alpha}/(\theta t)^{2\beta}\right] ,
 \label{eqcov}
\end{equation}
where $g(u)$ is an universal scaling function, and $A_h$ is the amplitude associated to the height difference correlation function $C_h$ (eq. \ref{corr}), i.e. $C_h(l) = A_hl^{2\alpha}$ over the $l_g \ll l \ll \xi$ scale. Functions $g(u)$ are analytically known for different initial conditions in 1D \cite{Covariance}, and have been experimentally confirmed in turbulent growth of nematic liquid-crystal phases \cite{Takeuchi11}. Universal scaling functions also exist in 2D, as observed numerically \cite{Healy14,Ismael14} and confirmed in the growth of oligomer films \cite{Healy14}. 

Noting that $C_s(0,t) = \left\langle h^2 \right\rangle_c$, from eq. \ref{eqansatz} one has that $g(0)=C_s(0,t)/(\theta t)^{2\beta} = \left\langle \chi^2 \right\rangle_c$. The \textit{universal} value of $\left\langle \chi^2 \right\rangle_c$ is numerically known to be $\left\langle \chi^2 \right\rangle_c \simeq 0.24$ \cite{Healy12,Ismael14}, for 2D-KPZ interfaces with flat IC. We estimate the experimental value of $\theta$ by assuming the universality of $\left\langle \chi^2 \right\rangle_c$ in $\theta \simeq \left[\left\langle h^2 \right\rangle_c/\left\langle \chi^2 \right\rangle_c\right]^{\frac{1}{2\beta}}/t$, and using our estimates for $\left\langle h^2 \right\rangle_c=w^2$ from Fig. \ref{fig1}(d). We obtain $\theta \approx 24\times 10^3$ $[nm^{1/\beta}]/[$min$]$ for $t=2880$ min and $\theta \approx 7\times 10^3$ $[nm^{1/\beta}]/[$min$]$ for $t=5760$ min. The different values are expected due to statistical fluctuations in $w$ [Fig. \ref{fig1}(d)]. 

\begin{figure}[h]
\centering
\includegraphics[width=11cm]{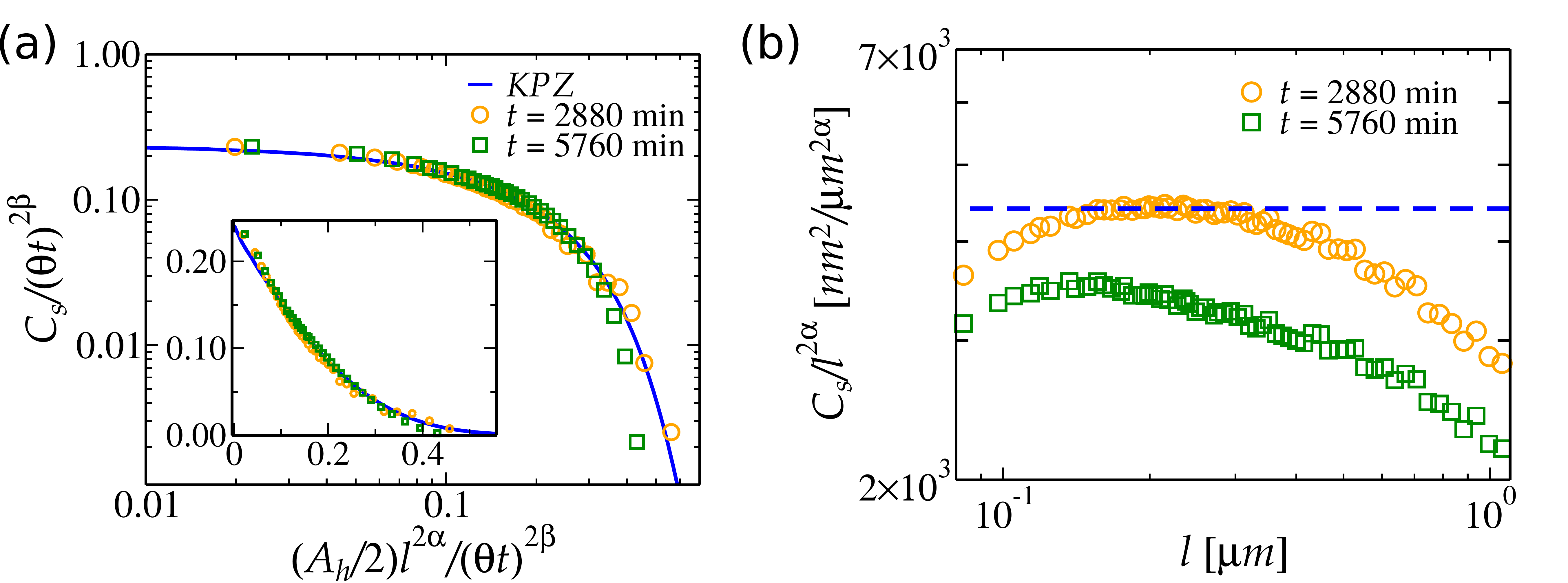}
\caption{(a) Rescaled spatial covariance for KPZ class \cite{Ismael14} (solid line) compared with the ones for CdTe surfaces (symbols). Inset shows the same data in linear scale. (b) Rescaled height difference correlation function $C_h/l^{2\alpha}$ versus $l$ for CdTe surfaces at long times. The dashed line indicates the estimate of $A_h$ for $t = 2880$ min. Rescaling were performed using $\alpha=0.3869$ \cite{Pagnani16} and $\beta=0.241$ \cite{Odor16}.}
\label{fig5}
\end{figure}

Now, considering the obtained values for $\theta$, we  find $A_h$ by making the rescaled $C_s(l,t)$ curves to collapse onto the KPZ universal scaling function $g(u)$ [Fig. \ref{fig5}(a)]. Such procedure yields $A_h \approx 5000$ $[nm^2/\mu m^{2\alpha}]$ for $t=2880$ min and  $A_h \approx 4400$ $[nm^2/\mu m^{2\alpha}]$ for $t=5760$ min. On the other hand, from the definition of $A_h$, we also estimate its value from the (expected) plateaus in the $C_h(l)/l^{2\alpha} \times l$ curves, as done in Fig. \ref{fig5}(b). Data for $t = 2880$ min has a clear plateau at $A_h \approx 4400$, which is quite close to the value estimated from the collapse of $C_s$. For $t = 5760$ min, however, the plateau is absent due to an effective $\alpha$ exponent a bit smaller than the KPZ one in the $C_h \times l$ scaling [Fig. \ref{fig1}(e)].

Finally, knowing $\theta$ and $A_h$, we determine the experimental value of $\lambda$ - the key KPZ parameter. Using the relation $A_h/A \approx 0.6460$ \cite{Healy14} and the $A_h$ from the collapses, we obtain $A \approx 7740$ $[nm^2/\mu m^{2\alpha}]$ for $t=2880$ min and $A \approx 6656$ $[nm^2/\mu m^{2\alpha}]$ for $t=5760$ min. Re-inserting these values into the definition of $\theta$, we find $\lambda \approx 3.6 \times 10^{-2}$ $nm/s$ and $\lambda \approx 1.5 \times 10^{-2}$ $nm/s$, respectively, for $t=2880$ min and $t=5760$ min. These values are rather small when compared to that reported for oligomer films ($\lambda \approx 5$ $nm/s$) \cite{Healy14}, but they are consistent with the CdTe growth rate $F \approx 23 \times 10^{-2}$ $nm/s$. Moreover, a small $\lambda$ is consistent with the very slow PSS-KPZ crossover found here. This may also explain the initial uncorrelated growth observed in CdTe/Si(100) \cite{Renan15} for $T = 150\,^{\circ}\mathrm{C}$, which possibly gives place to KPZ scaling at longer times. We remark that at higher temperatures, larger values of $|\lambda|$ are expected, as discussed in \cite{Renan15}.

It is noteworthy the good data collapse between covariances for CdTe surfaces and 2D-KPZ models [Fig. \ref{fig5}(a)]. This collapse provides additional evidence of an universal 2D-KPZ spatial covariance and, at the same time, gives a final confirmation that CdTe/Kapton films evolves asymptotically according to the KPZ class.


\section*{Summary and final discussions}

We have analyzed the kinetic roughening of CdTe films deposited on Kapton substrates at relatively low temperature and intermediate deposition rate. The rough and non-crystalline (polymeric) substrate allows, at the submonolayer regime, the formation of small 3D-CdTe grains with several crystallographic orientations and large height differences, yielding a large global roughness at very initial growth times. Smoothening and a long pseudo-steady state (PSS) regime take place in the sequence, as a consequence of grain collisions, with the former occurring at the transition between the submonolayer and multilayer growth. In the PSS regime, the smoothening comes from the collisions of high [mostly (111)] grains after covering up smaller [mostly non-(111)] ones. Interesting, in such regime roughening and smoothening are counter balanced, so that interface width and correlation length $\xi$ do not grow, exactly as they would do in a genuine steady state of finite-size systems. The reasoning of a combined surface roughening and smoothening leading to a PSS is supported by simulations of a 1D phenomenological model, which captures main features of grain coalescence in CdTe growth. On the other hand, the asymptotic CdTe growth regime is shown to belong to the KPZ class by $i)$ several (independently measured) scaling exponents, in addition to universal $ii)$ height, $iii)$ local square-width and $iv)$ extreme height distributions. Final striking evidence of KPZ scaling is given by $v)$ the spatial covariance, which allowed us to estimate the KPZ ``excess velocity'' of the studied films as $\lambda$ $\sim 10^{-2}$ $nm/s$. Overall, these results support that high vacuum vapor deposition of polycrystalline CdTe is a standard system belonging to KPZ class, regardless the substrate nature. This might stimulate previous works on CdTe grown on glass \cite{Sukarno06, Silvio11} to be revisited considering longer deposition times. Of course, for CdTe deposited under other growth conditions, e.g. by sputtering \cite{Amar14}, where inhomogeneous flux of particles and shadow effects can play a role, KPZ scaling may not take place.

For width fluctuations in the limit of low correlations ($\xi \ll l$), we have experimentally demonstrated the emergence of log-normal distributions, in both PSS and KPZ regimes, confirming recent conjectures \cite{Ismael15}. Moreover, from the temporal scaling of their cumulants, we found scaling exponents in good agreement with the KPZ ones - also in agreement with those obtained by the standard width and spatial correlation function. This fact demonstrates that the generalized Family-Vicsek scaling \cite{Ismael15} is an useful tool even for estimating exponents from experiments.

We remark that the PSS regime could be easily misunderstood as an Edwards-Wilkinson growth, since both are characterized by vanishing $\beta$ and Gaussian height distributions. This shows the necessity of measuring as many quantities as possible to confirm the class of real growing systems, specially not relying only upon the exponents. Note that we had to grow films up to $t = 5760$ min (4 days!) in order to observe the KPZ scaling regime only one decade in time (in $w \times t$ plot). Similar difficulty was already observed in previous KPZ systems \cite{Cuerno00}. The presence of grainy morphology also severely hampers the calculation of the exponent $\alpha$. A consequence of using only the exponents from the traditional Family-Vicsek scaling, usually estimated from short deposition times, is the large number of existing works reporting scaling analyses of real interfaces without association to any universality class.

We also point that an initial PSS regime might be rather general, since it was also observed in organosilicone films deposited by chemical vapor deposition at atmospheric pressure on polymeric substrates \cite{Premkumar12,Lu16} and in plasma etched polymeric films \cite{Bae15}. Even though the microscopic origin of this behavior seems quite different in such systems, it is somewhat intriguing that all these evidences of the PSS are associated with polymeric substrates. 

Finally, the KPZ mechanism in CdTe films comes from the packing of grains, which yields a velocity excess in the growth, as explained in \cite{Renan14,Renan15}. Obviously, results presented here do not have any relation with a possible artificial KPZ scaling induced by AFM tip effects \cite{Sidiney}. Note that typical tip radius ($r$) is $r \approx 10-20$ $nm$, corresponding to the size of our image pixel ($a \approx 19.5$ $nm$), while the average CdTe grain size is $l_g \approx 0.7$ $\mu m$ in the asymptotic KPZ regime. Namely, $l_g$ is at least 35 times larger than $r$ and, in such situation, possible effects of the AFM tip are negligible. More important, even when CdTe grains are smaller and some AFM tip effect could be expected, KPZ scaling is not found neither here (at short times) nor elsewhere \cite{Sukarno06, Silvio11}.

\section*{Acknowledgements}

RALA thanks fruitful discussions with I. S. S. Carrasco, F. A. A. Reis and T. Sasamoto, and a careful reading of preliminary versions of the paper by I. S. S. Carrasco and K. A. Takeuchi. Authors acknowledge support from Brazilian funding agencies CAPES, FAPEMIG and CNPq. RALA acknowledge financial support by KAKENHI,JSPS No16J06923.

\section*{Author contributions statement}

RALA grew samples, performed AFM measurements, built computational algorithms for data analyzes and wrote the paper. SOF designed, managed the whole experiment and analyzed the results. I. Ferraz also grew films and contributed with AFM and XRD measurements. TJO analyzed the results, performed simulations and wrote the paper. All authors reviewed the manuscript.

\newpage

{\centering {\huge {\bf Supplementary Information}}}

\maketitle

\section{X-Ray diffraction measurements}

\begin{figure}[!h]
\centering
\includegraphics[width=7.cm]{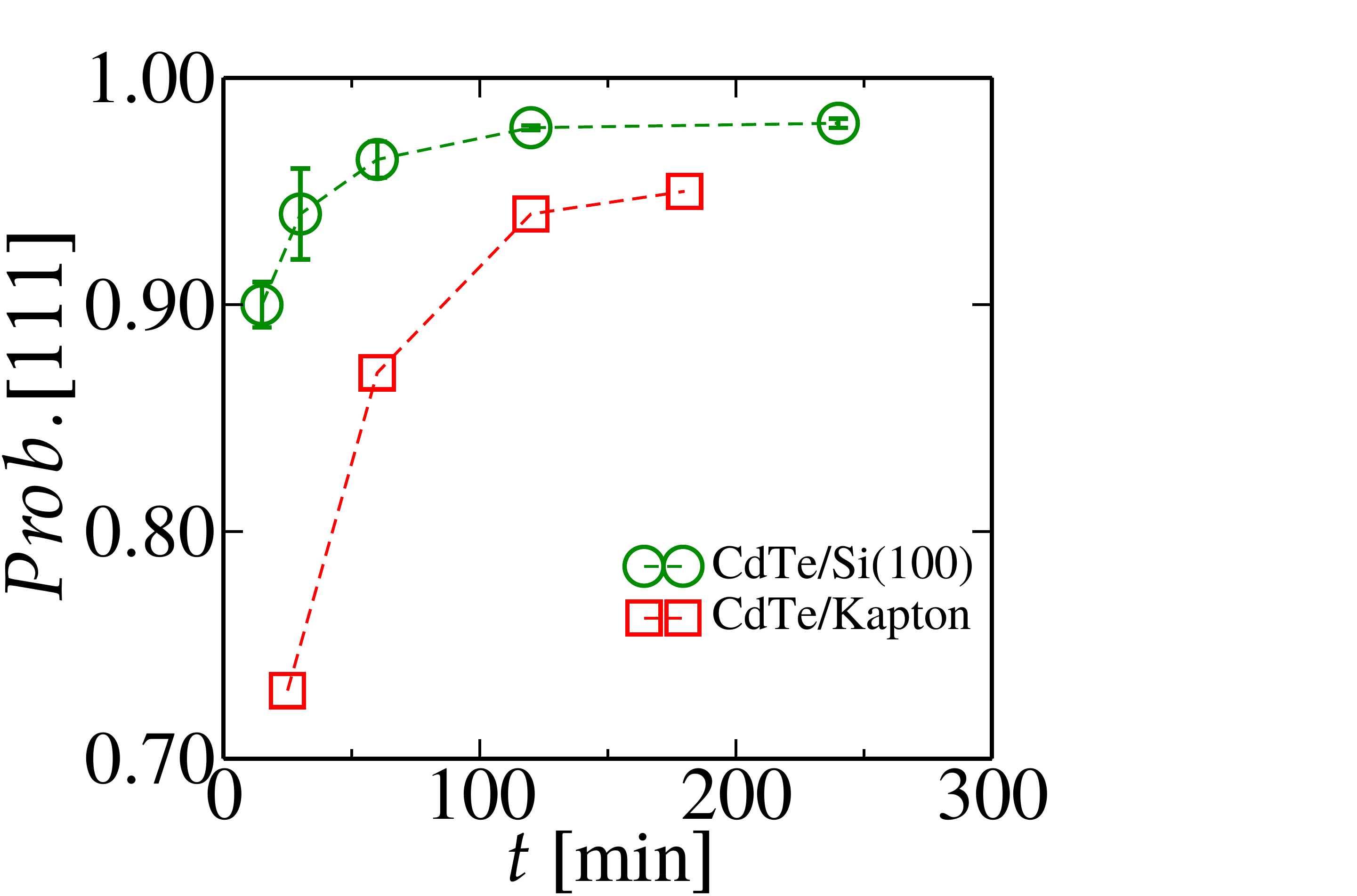}
\caption{Probability of finding a CdTe grain in the [111] direction as a function of time. Data for CdTe/Si(100) is the same reported in \cite{Renan15}.}
\label{Sfig1}
\end{figure}

We used a D8-Discover (BRUKER) X-Ray diffractometer (XRD) in the $\theta-2\theta$ coupled mode (radiation $\lambda_{CuK\alpha} = 0.15418$ nm) for studying CdTe crystalline structure as deposition proceeds. From the XRD spectra (not shown), we calculated the probability of finding grains in the [111] direction, relative to all other allowed directions, as:
\begin{equation}
P([111],t) = \frac{I_{111}(t)/A_{\theta-2\theta(\theta_{111})}}{\sum_{hkl} I_{hkl}(t)/A_{\theta-2\theta}(\theta_{hkl})}
\end{equation}
where $I_{hkl}$ is the intensity of the peak $hkl$, and $A_{\theta-2\theta}(\theta_{hkl})$ is the absorption factor for the $\theta-2\theta$ geometry, dependent on $\theta_{hkl}$ angle \cite{Birkholz}. Figure S\ref{Sfig1} shows the temporal evolution of probabilities $P([111], t)$ for both CdTe grown on Kapton and on Si(001) substrates at same conditions. On both substrates, there is an increasing of $P([111],t)$ with $t$, which indicates the existence of a preferential growth (i.e. a texture) in $[111]$ direction. However, the curves are rather different for short deposition times, with $P([111], t)$ significantly smaller for Kapton substrates. In short, (111)-grains take longer time to dominate the film structure in the Kapton case, probably due to the non-crystalline nature of the polymer. As discussed in the main text, this difference is the key to explain the existence of the pseudo-steady state (PSS) regime for Kapton.

\section{Additional data for width and extremal height distributions}

\begin{figure}[!t]
\centering
\includegraphics[width=14.5cm]{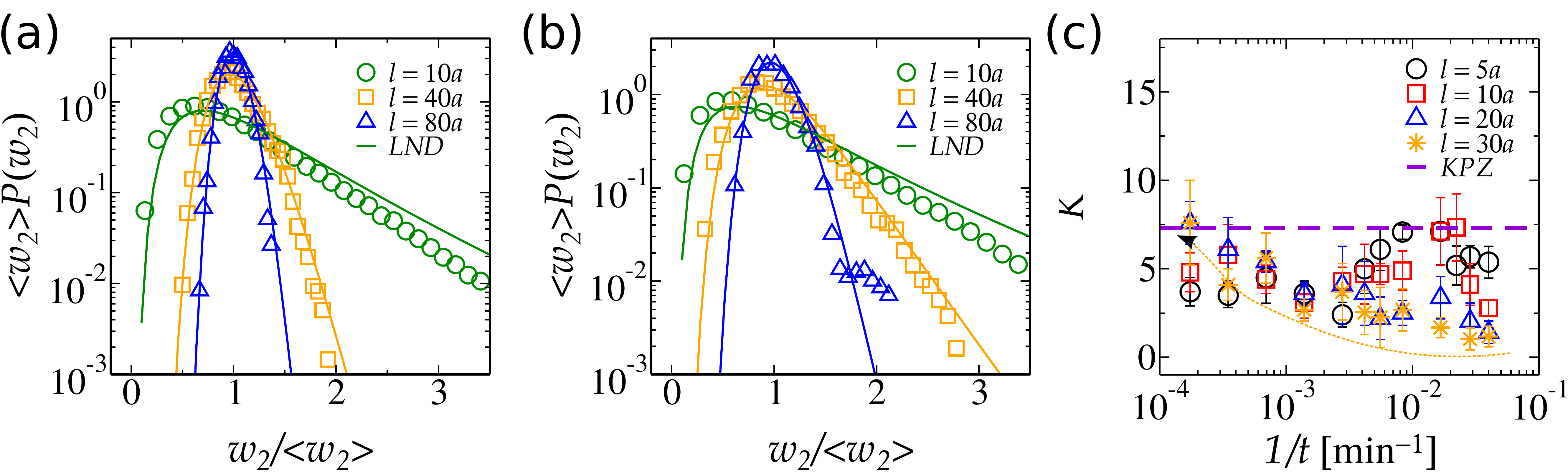}
\caption{Rescaled width distributions $P(w_2)$ (symbols) for CdTe surfaces for (a) $t = 120$ min (PSS regime) and (b) $t = 720$ min (KPZ regime). Solid lines are LNDs with the respective (experimentally calculated) parameters $\mu$ and $\sigma$. (c) Kurtosis $K$ of $P(w_2)$ as a function of $1/t$ for several window sizes.}
\label{Sfig2}
\end{figure}

In Fig. 3(d) of the main text we compare width distributions (rescaled to zero mean and unitary variance) with the log-normal distribution (LND). A more complete analysis can be done by comparing distributions rescaled to unitary mean $P(x)$, with $x \equiv \frac{w_2}{\langle w_2 \rangle}$, which are expected \cite{Ismael15} to be given by the LND:
\begin{equation}
 P(x,t) = \frac{1}{\sqrt{2\pi}\sigma x}\exp{\left\lbrace -\frac{\left[ \ln(x) - \mu \right]^2 }{2\sigma^2}\right\rbrace},
\end{equation}
 when the correlation length $\xi$ is much smaller than the window size $l$ ($\xi \ll l$). In the equation above, $\mu(t) \equiv \langle \ln(x) \rangle$ and $\sigma(t) \equiv \sqrt{\langle \ln^2(x) \rangle - \langle \ln(x) \rangle^2}$. We rescaled distributions in such way, for fixed times in PSS and KPZ regimes and different $l$'s [see Figs. S\ref{Sfig2}(a) and S\ref{Sfig2}(b)]. Indeed, in both regimes we observe a nice agreement between the width distributions and their respective LNDs (with $\mu$ and $\sigma$ obtained directly from the experimental data). As expected, as larger $l$ is, narrower the distributions are, since they are converging to a delta function. Moreover, in the limit of small $l$ the collapses worsen, since the condition $\xi \ll l$ is not fully satisfied.
 
The convergence of the kurtosis $K$ for the stationary width distribution (for $\xi \gg l$) is presented in Fig. S\ref{Sfig2}(c). Similarly to the cumulant ratios $R$ and $S$ (shown in Figs. 4(a) and 4(b) of the main text, respectively), we observe that $K$ converges to the KPZ value as $t$ and $l$ becomes large (with $\xi \gg l$).

\begin{figure}[h]
\centering
\includegraphics[width=14.5cm]{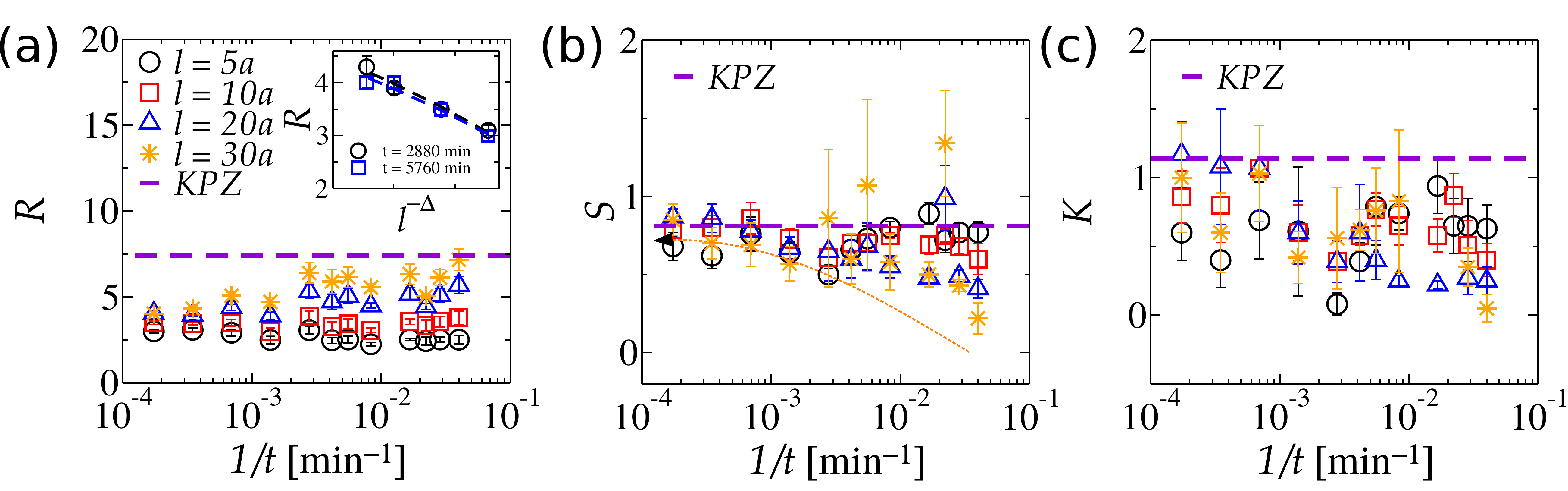}
\caption{Cumulant ratios (a) $R$, (b) skewness $S$ and (c) kurtosis $K$ for maximal height distributions as functions of $1/t$, for several window sizes ($l \ll \xi$). The inset in (a) shows an extrapolation of $R$ as $l^{-\Delta}$ goes to zero, where $\Delta = 0.17$ is used to linearize the data.}
\label{Sfig3}
\end{figure}

Figure S\ref{Sfig3} shows cumulant ratios for (local) maximal height distributions $P(m)$ calculated in the stationary regime ($\xi \gg l$). The skewness is in striking agreement with the value expected \cite{Ismael15} for KPZ class, presenting small time- and $l$-corrections. Values of $K$ for large $t$ and $l$ also agree quite well with the KPZ one. On the other hand, the ratio $R \equiv \langle m \rangle_c/\langle m^2\rangle_c^{1/2}$ is still rather different from the KPZ value, even for large $t$ and $l$. One way for dealing with this slow convergence is to take the $R$ values for the longest times and extrapolate them for $l \rightarrow \infty$. Performing such procedure [see the inset of Fig. S\ref{Sfig3}(a)], we find $R = 7.3(5)$, which is basically the same $R$ value obtained from simulations of KPZ models \cite{Ismael15}.

\end{document}